\documentclass{aa}  

\usepackage{graphicx}
\usepackage{txfonts}
\usepackage[colorlinks=true, allcolors=blue]{hyperref}
\begin{document}

   \title{Expanding the inventory of spectral lines used to trace atmospheric escape in exoplanets}

   \author{D.C. Linssen
          \thanks{E-mail: d.c.linssen@uva.nl}
          \and
          A. Oklop\v{c}i\'{c}
          }

   \institute{Anton Pannekoek Institute for Astronomy, University of Amsterdam,
              Science Park 904, 1098 XH Amsterdam, The Netherlands
             }

   \date{Received 3 April 2023 / Accepted 12 June 2023}
 
  \abstract
    {Escaping exoplanet atmospheres have been observed as deep transit signatures in a few specific spectral lines. Detections have been made in the hydrogen Ly-$\alpha$ line, the metastable helium line at 10830~Å and some UV lines of metallic species. Observational challenges, unexpected non-detections and model degeneracies have generally made it difficult to draw definitive conclusions about the escape process for individual planets. Expanding on the suite of spectral tracers used may help to mitigate these challenges. We present a new framework for modeling the transmission spectrum of hydrodynamically escaping atmospheres. We predict FUV to NIR spectra for systems with different planet and stellar types and identify new lines that can potentially be used to study their upper atmospheres. Measuring the radius in the atmosphere at which the strongest lines form puts them into context within the upper atmospheric structure. Targeting a set of complementary spectral lines for the same planet will help us to better constrain the outflow properties.}

   \keywords{}

   \maketitle

\section{Introduction} \label{sec:intro}
Planets that orbit close to their host stars can experience extreme levels of high-energy stellar irradiation, which can drive hydrodynamic escape of their atmospheres \citep[see review by][]{owen_atmospheric_2019}. At the same time, heat released by the cooling core of a planet can also supply energy to drive the escape process \citep{ginzburg_core-powered_2018}. Hot Jupiters undergoing escape have been shown to be stable over their lifetimes \citep[e.g.][]{murray-clay_atmospheric_2009, vissapragada_upper_2022}. Less massive planets such as hot Neptunes might however lose a significant fraction of their atmosphere in this way \citep{ionov_survival_2018}, which can affect their composition and stability. This process has been invoked to explain the bimodal radius distribution of sub-Neptune planets around $\sim$1.6~$R_\oplus$ \citep[e.g.][]{owen_evaporation_2017, ginzburg_core-powered_2018}, as well as the lack of sub-Jovian planets at short orbital distances \citep[e.g.][]{owen_kepler_2013, lundkvist_hot_2016}. 

Atmospheric escape has been observed for a growing sample of exoplanets in a few specific spectral lines that trace the low-density gas in the outflow. Absorption in the wings of the Ly-$\alpha$ line has provided direct evidence of gas escaping beyond the planet's Hill radius \citep[e.g.][]{vidal-madjar_extended_2003}. This line allows us to study the regions where the planetary wind interacts with the stellar environment, but the wind-launching regions are obscured \citep{owen_fundamentals_2022}. The metastable helium triplet can be observed at high resolution in the near-infrared \citep[e.g.][]{nortmann_ground-based_2018, allart_spectrally_2018}. This line probes deeper into the planetary outflow than Ly-$\alpha$ and therefore does allow us to study the wind-launching mechanism \citep{oklopcic_new_2018}. For a small sample of exoplanets, strong absorption from metal species in the UV has also indicated the presence of a planetary outflow. This includes the Fe~II UV1 and UV2 multiplets \citep{sing_hubble_2019, cubillos_near-ultraviolet_2020}, the Mg~II doublet \citep{fossati_metals_2010, sing_hubble_2019}, the C~II~$\lambda$~1335 lines \citep{vidal-madjar_detection_2004, garcia_munoz_heavy_2021, ben-jaffel_signatures_2022}, the O~I~$\lambda$~1305 lines \citep{vidal-madjar_detection_2004, ben-jaffel_hubble_2013} and the Si~III~$\lambda$~1206 line \citep{linsky_observations_2010}.

Each of these spectral lines have their unique observational challenges. For example, the ISM and geocorona render the core of the Ly-$\alpha$ line inaccessible \citep[e.g.][]{lecavelier_des_etangs_evaporation_2010}, and the helium line preferentially forms in planets around K-type stars \citep{oklopcic_helium_2019}. Still, there are many exoplanets for which non-detections of atmospheric-escape tracers have been reported, even though atmospheric mass loss is expected based on e.g. the received irradiation (see \citealt{dos_santos_observations_2022} for an overview). Similarly for some planets, ongoing atmospheric mass loss has been observed in one or more spectral tracers, but has eluded detection in others. For example, in HD~189733~b, Ly-$\alpha$ \citep{lecavelier_des_etangs_evaporation_2010}, metastable helium \citep{salz_detection_2018} and O~I \citep{ben-jaffel_hubble_2013} have been detected, but Mg~II absorption has not \citep{cubillos_hubble_2023}. The interpretation of (non-)detected signatures has also not been straightforward. For example, different types of models have been used to interpret helium observations (e.g. a particle code as employed by \citet{allart_spectrally_2018}, a 3D hydrodynamics code as employed by \citet{wang_metastable_2021} and a 1D isothermal Parker wind model as employed by \citet{mansfield_detection_2018}), with varying predictions and inferences about some of the core properties of the outflow, such as the mass-loss rate. 

Finding new spectral lines that probe escaping atmospheres may mitigate these challenges and help to enhance our understanding of atmospheric mass loss. A step in this direction was taken by \citet{turner_investigation_2016}, who predicted spectral signatures for a uniform slab of material at conditions similar to that of escaping planetary gas. In this work, we present a new framework for producing synthetic transmission spectra of evaporating exoplanet atmospheres from FUV to NIR wavelengths and we use it to identify new spectral lines that can potentially be used to study atmospheric escape. We measure the atmospheric radii probed by the most prominent spectral lines in order to create a more complete picture of the upper atmosphere.

\section{Methods} \label{sec:methods}
\subsection{Modeling the escaping atmosphere with \texttt{Cloudy}} \label{sec:modeling_with_cloudy}

We model the escaping planet atmosphere as a Parker wind \citep{parker_dynamics_1958, lamers_introduction_1999} made with the \texttt{p\_winds} code \citep{dos_santos_p-winds_2022}. This is a parametrized outflow model with an isothermal temperature and mass-loss rate as parameters, as well as a chemical composition that we keep fixed at 90\% hydrogen and 10\% helium by number. We employ our iterative algorithm presented in \citet{linssen_constraining_2022} to simulate the Parker wind density profile with \texttt{Cloudy v17.02}\footnote{https://gitlab.nublado.org/cloudy/cloudy/-/wikis/home} \citep{ferland_cloudy_1998, ferland_2017_2017}. In short, our algorithm solves for a (non-isothermal) temperature structure based on the radiative heating and cooling rates as calculated by \texttt{Cloudy}, while also including the hydrodynamic thermal effects of expansion cooling and heat advection. Since the mass-loss rate and constant temperature are free parameters of the Parker wind model, many different models can be constructed for a given planet. The chosen temperature may however be unrealistic based on the stellar flux and density and velocity of the outflow. We leverage this to identify `self-consistent' Parker wind profiles through a comparison between the chosen isothermal temperature and \texttt{Cloudy}'s converged temperature structure. There is freedom in how to practically do this comparison and the most useful way depends on the aim of the modeling effort. In \citet{linssen_constraining_2022} we interpreted metastable helium detections and hence constructed a characteristic temperature in the helium line-forming region from the \texttt{Cloudy} simulations, which we compared to the original isothermal temperature. In this work we aim to simulate a whole suite of spectral lines probing different parts of the upper atmosphere. It is thus impossible to find one characteristic temperature for the line-forming regions of all these lines, and hence we choose to simply calculate the radial average of \texttt{Cloudy}'s temperature profile between 1 and 2 planetary radii and compare it to the isothermal value. At those radii the planetary wind is launched and respecting self-consistency in this region generally results in a selection of Parker wind models whose density and velocity profiles are most similar to self-consistent outflow models (see also Sec. \ref{sec:ates}). In this way we focus on obtaining a sensible overall outflow structure instead of focusing on one part of the outflow where a particular spectral line forms. Still, we reiterate that the choice is ultimately arbitrary. After checking the self-consistency of the models, for one value of the mass-loss rate we tend to find one Parker wind model with a self-consistent temperature and consider just this model. We thus effectively reduce the free parameters of the Parker wind model to only the mass-loss rate (and the composition). In Sec. \ref{sec:results} we explain which specific Parker wind model parameters we use in this work.

Once we have found a self-consistent Parker wind model, we calculate its transmission spectrum. We do not use \texttt{Cloudy}'s own predicted spectrum for this, but take its thermal and chemical atmospheric profiles and post-process these to calculate the spectral lines found in the NIST database\footnote{https://www.nist.gov/pml/atomic-spectra-database}. Appendix \ref{app:RT} describes the procedure in more detail.

\subsection{Selecting suitable spectral lines} \label{sec:selecting}

We calculate mid-transit transmission spectra over the wavelength range 911-20,000~Å (see Fig. \ref{fig:209_spectrum} for an example spectrum). Below 911~Å, the continuum opacity of hydrogen (and helium) ionization renders other spectral lines inaccessible. Above 20,000~Å, there are almost no strong atomic spectral lines. In each  transmission spectrum, we then search for spectral lines that are strong enough to be potentially observable. For every atomic and ionic species, we find the strongest spectral line in the UV (<4000~Å) and the strongest line in the VIS/NIR (>4000~Å). If the strongest UV line has more than 1\% excess absorption at line center, we include it in our sample of suitable tracers, and then do the same for the strongest VIS/NIR spectral line if it is stronger than 0.1\%. We chose to split the wavelength domain in two and use different thresholds because the UV and VIS/NIR are typically observed by different instruments with different precision. For some species such as e.g. Fe$^{+}$, there are many strong spectral lines that could be used as upper atmospheric tracers (and it may be possible to use cross-correlation techniques), but we include here only the strongest spectral line to keep the analysis clear and straightforward. We emphasize that we are modeling hydrodynamic upper atmospheres that are completely atomic/ionic, and thus any molecular absorption bands originating from the `lower' atmosphere are absent from our spectra.

\subsection{Calculating the radius where spectral lines form} \label{sec:lfa}

From looking at the transmission spectrum, one might naively assume that deeper spectral lines originate at higher altitudes in the atmosphere. While this is true in many cases, it need not always be, as a spectral line of a given depth might originate from a confined region close to the planet with relatively high optical depth or from a more diffuse, extended region with relatively low optical depth, thereby probing a different part of the outflow. In this work, we want to distinguish between these cases and measure the radius of the atmospheric region in which each spectral line forms. 


   \begin{figure}
   \centering
   \includegraphics[width=\hsize]{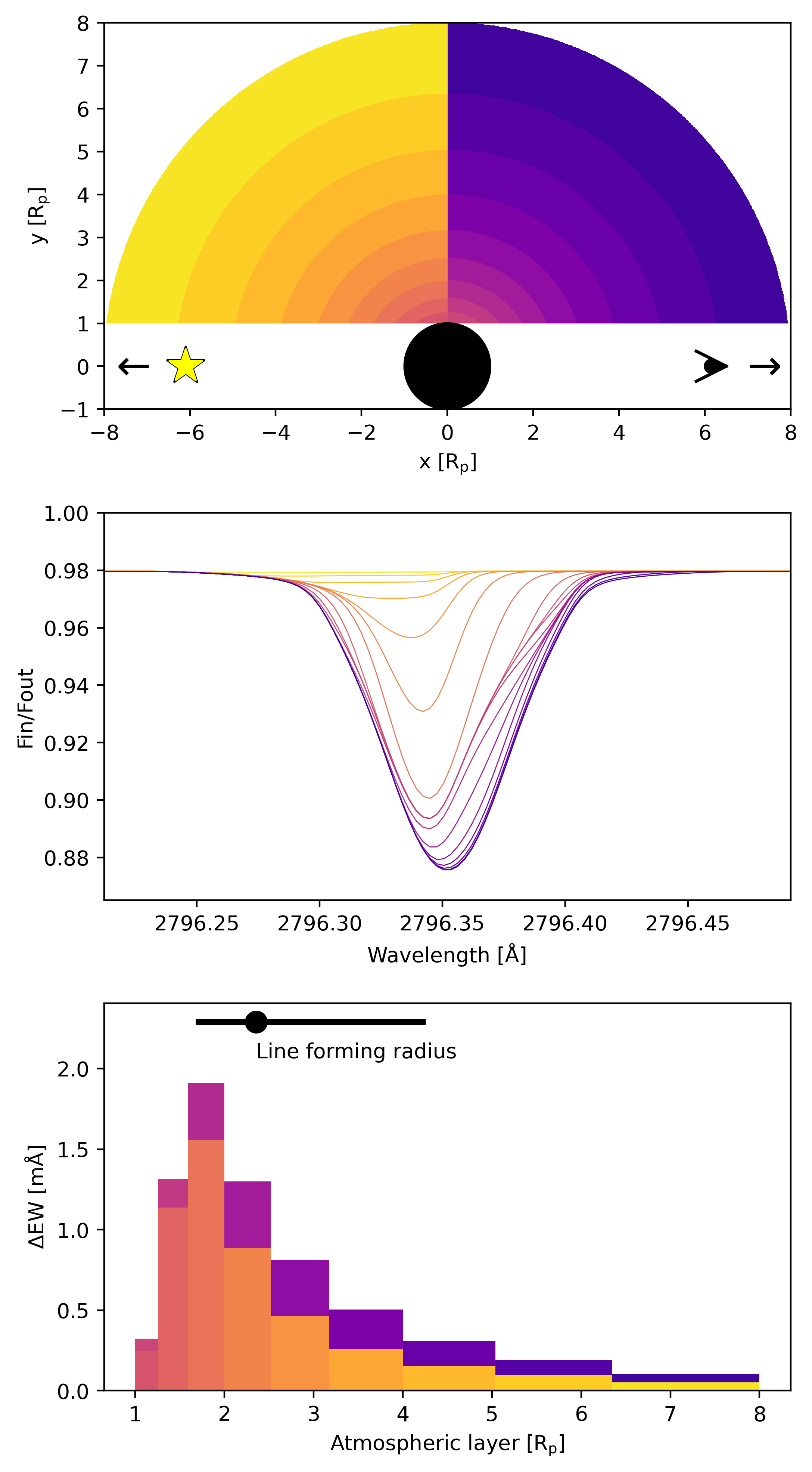}
      \caption{Illustration of how the line-forming radius is defined for the Mg II line at 2796~Å. We consecutively keep adding deeper atmospheric layers and track the increase in equivalent width (EW) of the spectral line. Top panel: 2D projection of the atmosphere separated into different layers (indicated by different colors). We show 18 layers for clarity, but our real calculations consist of 60 layers. The planet is indicated by the black circle. The star is to the left and the observer to the right. Middle panel: the `observed' spectral line when only including the part of the atmosphere with the corresponding color as well as lighter shaded layers (i.e. more yellow layers). Bottom panel: contributions to the EW of the line from each atmospheric layer. The black point indicates the line-forming radius and standard deviations that are calculated based on the histogram and defined in Eqs. \ref{eq:lfa} and \ref{eq:sig_lfa} (the plotted $\Delta$EW value of the black point is meaningless).}
         \label{fig:lfa_EW_definition}
   \end{figure}
   
To this end, we devise a measure for the `line-formation radius' of a given spectral line, based on the contribution to the equivalent width (EW) of the line of each atmospheric layer (illustrated in Fig. \ref{fig:lfa_EW_definition}). We divide our atmosphere into a dayside ($x<0$) and nightside ($x>0)$ and further into layers logarithmically spaced in radius. We then start by calculating the transit spectrum and its EW while only including the material of the uppermost layer on the dayside (the yellow layer between 8$R_p$ and 6.3$R_p$ in the figure). We then add a deeper layer to the atmosphere (the orange layer between 6.3$R_p$ and 5$R_p$ in the figure) and re-calculate the transit spectrum and its EW. The increase in EW is the contribution of the second layer, and these contributions per layer are shown in the bottom panel of Fig. \ref{fig:lfa_EW_definition}. We repeat this process until all dayside layers are included, after which we add the lowest nightside layer (the pink layer between $1R_p$ and $\sim$1.3$R_p$ in the figure) and work our way outwards again. We finally co-add the EW contributions of the dayside and nightside layers of equal radius, which results in a histogram like the bottom panel of Fig. \ref{fig:lfa_EW_definition}. 

The reason we treat the dayside and nightside separately is that this is also the order in which a light ray originating from the star would probe the planetary atmosphere. This is important when the dayside layer at, for example, $8R_p$ absorbs the first light and thus contributes to the EW, but then the atmosphere becomes optically thick at lower altitudes after which the nightside layer at $8R_p$ does not contribute significantly to the EW anymore. If we instead had started by calculating the EW using the uppermost layer of both day- and nightside, the nightside would have contributed to the EW as the gas would appear to be optically thin still.

From the histogram, we finally define the line-forming radius $r_{line}$ as the average of the radial bins $r$ weighted by their EW contribution ($\Delta$EW):
\begin{equation} \label{eq:lfa}
    r_{line} = \frac{\sum_r r \cdot \Delta\mathrm{EW}(r)}{\sum_r \Delta\mathrm{EW}(r)}
\end{equation}
and define a corresponding standard deviation that gives an indication of the spread of radii probed
\begin{equation} \label{eq:sig_lfa}
    \sigma_r = \sqrt{\frac{\sum_{r'} (r'-r_{line})^2 \cdot \Delta\mathrm{EW}(r')}{\sum_{r'} \Delta\mathrm{EW}(r')}}
\end{equation}
which we do separately for radii above and below $r_{line}$ (i.e. $r' > r_{line}$ and $r'<r_{line}$, respectively). Performing this exercise for each identified strong spectral line allows us to make a plot that resembles a spectrum but presents the radii at which the lines form instead of the transit depth.

\section{Results} \label{sec:results}
\subsection{Model grid} \label{sec:model_grid}
\begin{figure*}
   \centering
   \includegraphics[width=\hsize]{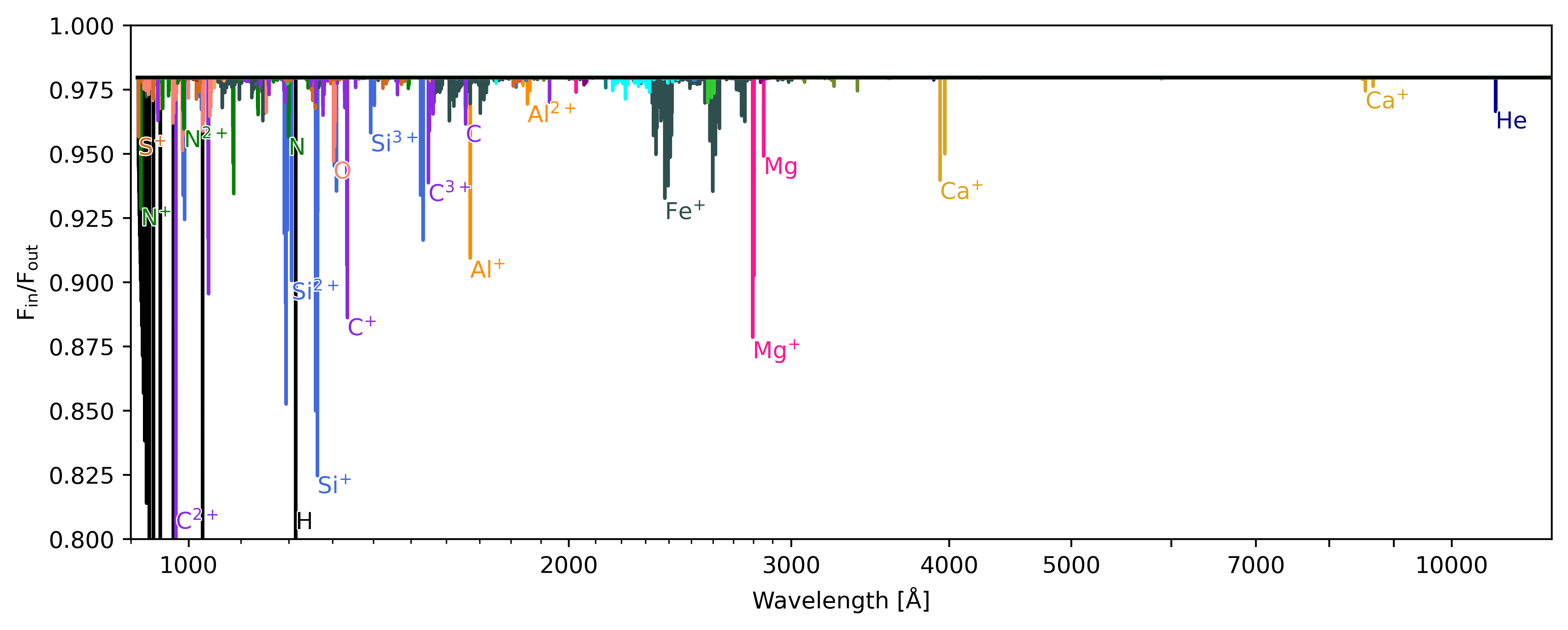}
      \caption{Transmission spectrum of the hot Jupiter planet around a G2 star, for a Parker wind profile with $\dot{M}=10^{10}$~g~s$^{-1}$ and $T=8000$~K. The strongest line of each atomic/ionic species is marked if it exceeds our thresholds of 1\% at $\lambda < 4000$~Å and 0.1\% at $\lambda>4000$~Å and its line-forming radius is shown in the top right panel of Fig. \ref{fig:209_10}.}
         \label{fig:209_spectrum}
   \end{figure*}

\begin{figure*}
   \centering
   \includegraphics[width=\hsize]{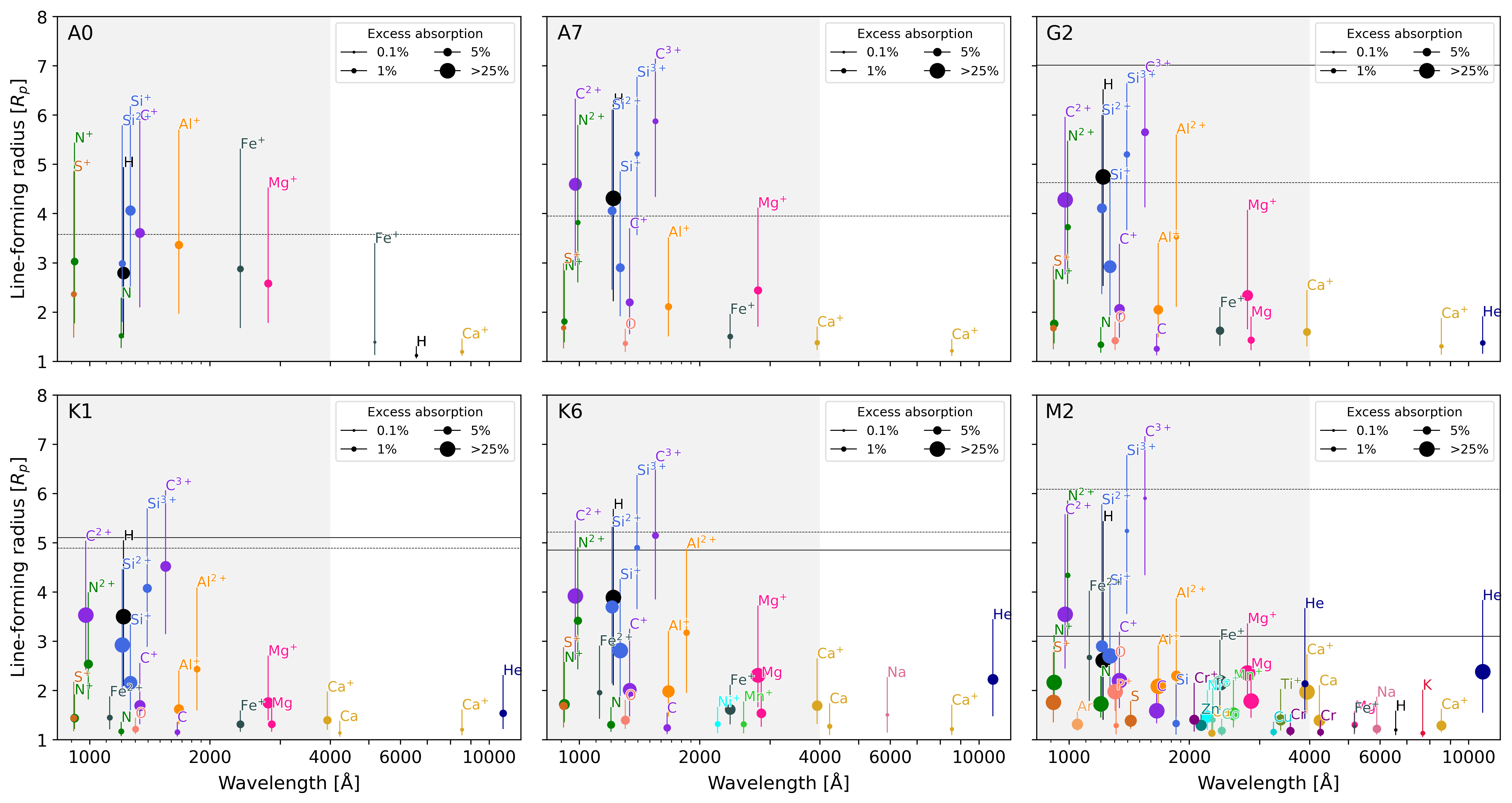}
      \caption{Line-forming radii for the hot Jupiter planet around stars of different spectral types. The outflow models used are Parker winds with $\dot{M}=10^{10}$~g/s and (self-consistent) temperatures of 7000, 7500, 8000, 8000, 8000, 8000 K for the A0, A7, G2, K1, K6, M2 spectral types, respectively. The marker size indicates the excess absorption at line center. The figures show all spectral lines deeper than 1\% at $\lambda<4000$~Å and deeper than 0.1\% at $\lambda>4000$~Å (Table \ref{tab:lines} lists the wavelength of each line). In every panel, the gray shaded area demarcates the 4000~Å boundary, the dotted line indicates the Hill radius and the solid line shows the stellar radius. Gas at larger radii than the stellar radius can contribute to the transmission spectrum if it is located `along the direction of the stellar light rays', instead of at high impact parameter.}
         \label{fig:209_10}
   \end{figure*}

\begin{figure*}
   \centering
   \includegraphics[width=\hsize]{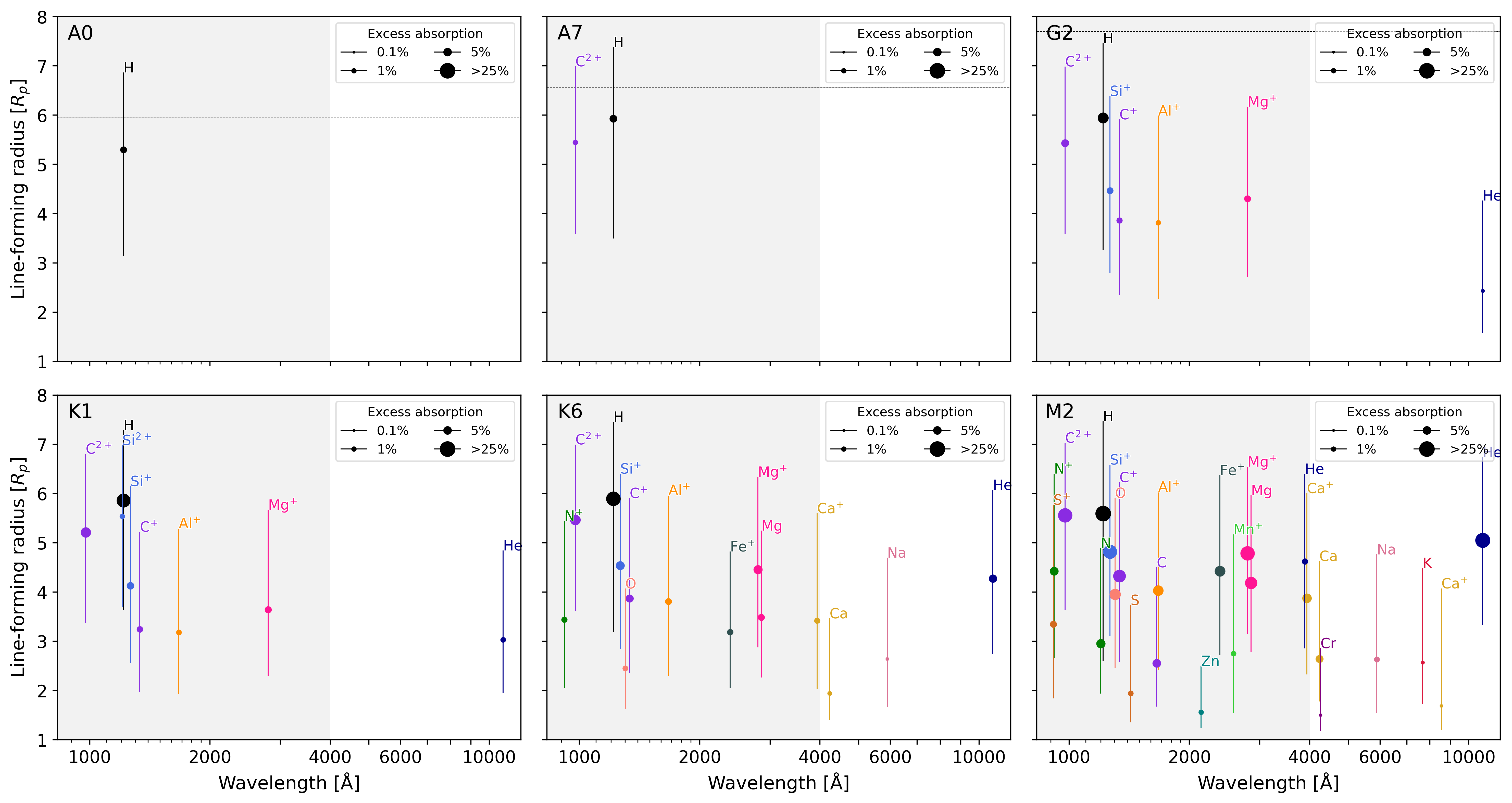}
      \caption{Similar to Fig. \ref{fig:209_10} but for the hot Neptune planet. The outflow models used are Parker winds with $\dot{M}=10^{10}$~g/s and (self-consistent) temperatures of 5000, 5000, 5500, 5000, 5000, 5500 K for the A0, A7, G2, K1, K6, M2 spectral types, respectively.}
         \label{fig:11_10}
   \end{figure*}

Our goal is to find the spectral lines that trace upper exoplanet atmospheres and assess where these spectral lines form. The results generally depend on the parameters of the system, hence we investigate a grid of systems with different stellar spectral types and planet gravities. In terms of the spectral type of the host star, we explore the A0 and A7 models described in Sec. 2.2 of \citet{oklopcic_helium_2019}, which are based on the work of \citet{fossati_extreme-ultraviolet_2018}, a solar (G2) spectrum from combined TIMED/SEE \citep{woods_solar_2005} and SORCE \citep{rottman_sorce_2005} data, and HD~97658 (K1), HD~85512 (K6) and GJ-832 (M2) spectra from the MUSCLES survey \citep{france_muscles_2016, youngblood_muscles_2016, p_loyd_muscles_2016}. The spectral energy distributions (SEDs) are shown in Fig. \ref{fig:SEDs}. In terms of the planet gravity, we explore a hot Jupiter planet with the mass and radius akin to HD~209458~b ($R=1.38R_J$, $M=0.73M_J$, $\phi=GM/R=10^{12.98}$~erg~g$^{-1}$) and a hot Neptune planet with the mass and radius akin to HAT-P-11~b ($R=0.389R_J$, $M=0.0736M_J$, $\phi=10^{12.53}$~erg~g$^{-1}$). We use Parker wind models with mass-loss rates of $\dot{M}=10^{10}$~g~s$^{-1}$ as in \citet{linssen_constraining_2022} we retrieved similar values when fitting a sample of metastable helium observations, but we do explore lower and higher mass-loss rates in Sec. \ref{sec:mdot}. The Parker wind temperatures are found following the methodology described in Sec. \ref{sec:modeling_with_cloudy} and are listed in the captions of the figures. We fix the semi-major axis of all models to $a =0.05$~AU (changing to $a=0.025$~AU only had a minor impact on the results). We keep the composition of the gas fixed at solar. Fig. \ref{fig:209_spectrum} shows the transmission spectrum of the hot Jupiter planet around a G2 star. The other transmission spectra are not shown in this work but are available for download\footnote{The transmission spectra and a reproduction package for all figures of this manuscript are available for download at \url{https://doi.org/10.5281/zenodo.8013349}}. Figs. \ref{fig:209_10} and \ref{fig:11_10} show the resulting line-forming radii. Table \ref{tab:lines} lists all unique spectral lines identified. Our model also predicts the hydrogen Ly-$\alpha$ line, but we caution that our model does not include interactions with the stellar wind which will shape the observable wings of the line. This makes the calculated Ly-$\alpha$ transit depth and formation radius unreliable.

\subsection{General trends with stellar spectral type}
We generally find more spectral lines that reach our excess absorption threshold around later type stars. This is in large part due to the transit geometry, since the 1\% and 0.1\% thresholds are reached more easily around late type stars as their radii are smaller. Fainter lines may however be easier to observe around earlier type stars as they have a higher luminosity and thus better signal-to-noise ratio. In terms of the different SEDs, the XUV flux differs by up to two orders of magnitude (even more for the A0 star) and the maximum is reached in the spectrum of the K1 star (Fig. \ref{fig:SEDs}). At higher XUV flux, the degree of ionization increases, and we tend to see stronger lines originating from (highly) ionized species and weaker lines originating from atomic species. For example, the C~IV~$\lambda$~1548 and Si~IV~$\lambda$~1393 lines are strongest around a K1 host star. The different XUV flux levels also result in slightly different self-consistent temperatures and mean molecular weight structures found for the Parker wind models. The A0 and M2 stars have the lowest XUV flux, resulting in lower hydrogen and helium ionization and hence higher $\mu$. At fixed mass-loss rate, a Parker wind model with lower $T/\mu$ has a denser outflow profile, and thus generally stronger lines.

\subsection{Dependence of specific lines on stellar spectral type}
In the VIS/NIR wavelength range, the strongest tracers are the metastable helium triplet and the calcium infrared triplet. The metastable helium line preferably forms around K-type stars as found in \citet{oklopcic_helium_2019} and corroborated by the current study. Fig. \ref{fig:209_10} shows that the transit depth of the calcium infrared triplet remains roughly constant with stellar spectral type, which means that the total amount of absorption must increase around earlier type stars (as their radii are larger). This can be explained by the fact that it originates from excited Ca$^+$ ions for which a main population channel is absorption of a $\sim$730~nm photon by the ground-state. Earlier type stars will have a higher flux of such NIR photons. This may make the Ca~II infrared triplet a highly complementary tracer to the metastable helium line by favoring `opposite' stellar types.

The H$\alpha$ line originates from hydrogen atoms that are excited (among other processes) through absorption of a stellar Ly-$\alpha$ photon. The A0 stellar SED template has the strongest Ly-$\alpha$ flux and this is why (for the hot Jupiter planet) we see stronger H$\alpha$ absorption there. Around the M2 star we see H$\alpha$ due to the favorable radius contrast as discussed above.

After Ly-$\alpha$, the strongest UV line is C~III~$\lambda$~977 around all stars except A0. Our models predict transit depths roughly half that of Ly-$\alpha$, yet this line has never been observed. Other lines that are strong irrespective of stellar type are Si~II~$\lambda$~1264, the Mg~II doublet, Si~III~$\lambda$~1206 and C~II~$\lambda$~1335. Both silicon lines have not been reliably detected (see Table \ref{tab:lines} and references therein).

\subsection{Relation between transit depth and line-forming radius}
Our results show that while there is certainly a trend in stronger spectral lines forming at larger radii, this is definitely not true for all lines. A good example is the N~III~$\lambda$~992 line, which forms at somewhat similar radii as the C~III~$\lambda$~977 line, yet is much weaker. For most lines, the radii probed are a lot larger than one would infer by doing a simple conversion of transit depth to apparent planet size. Such a conversion effectively assumes that the atmosphere is completely opaque up until said planet size but becomes completely transparent above it. Instead of this scenario, we find that for many lines there is important contribution from optically thin high altitude layers. Even though the density is low at these layers, noticeable absorption still occurs due to the total absorbing volume increasing with $r^3$ in a spherically symmetric atmosphere (we discuss this in more detail in Sec. \ref{sec:comp_domain}). Additionally, the degree of ionization increases with radius (see Fig. \ref{fig:ion_structure}) so that the relative (or even absolute) density of ion species rises with radius. This pushes their line-forming regions higher, but these lines are not necessarily strong (for example the C~IV line at 1548~Å).

\section{Discussion} \label{sec:discussion}
\subsection{Dependence on the mass-loss rate} \label{sec:mdot}
\begin{figure*}
   \centering
   \includegraphics[width=\hsize]{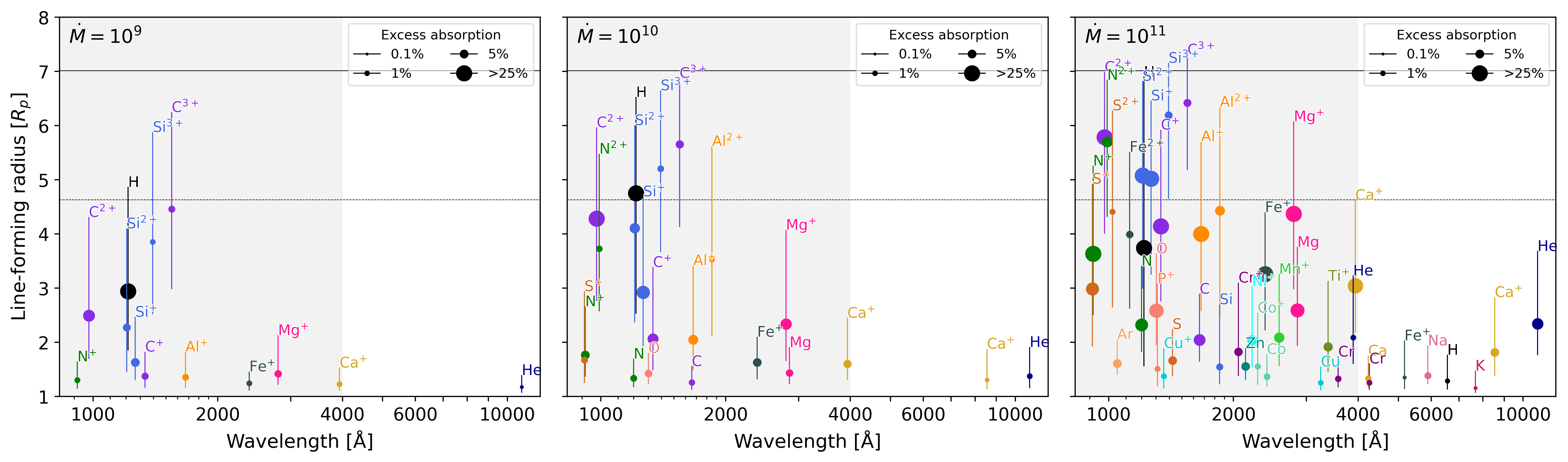}
      \caption{Line-forming radii for the hot Jupiter planet around a G2 star for different mass-loss rates. In the left panel we used $\dot{M}=10^9$~g~s$^{-1}$ and $T=7000$~K. In the middle panel we used $\dot{M}=10^{10}$~g~s$^{-1}$ and $T=8000$~K (identical to the upper right panel of Fig. \ref{fig:209_10}). In the right panel we used $\dot{M}=10^{11}$~g~s$^{-1}$ and $T=8000$~K.}
         \label{fig:209_comparison_Mdot}
   \end{figure*}

Some spectral lines that are generally thought to trace the upper atmosphere, such as e.g. H$\alpha$ and K~I~$\lambda$~7667 \citep[e.g.][]{keles_probing_2020, dos_santos_observations_2022}, are missing in many panels of Figs. \ref{fig:209_10} and \ref{fig:11_10}. The Parker wind profiles do not reach the column density needed for these lines to become stronger than 0.1\%. For the hot Jupiter planet around a G2 star, we also simulated a Parker wind model with $\dot{M}=10^{11}$~g~s$^{-1}$ and $T=8000$~K. As Fig. \ref{fig:209_comparison_Mdot} shows, we do see many more lines appearing when using a higher mass-loss rate, including the aforementioned lines in the optical wavelength range. Their line-forming radii indicate that most of these lines do in fact trace the upper atmosphere, but their absence from the sample at lower mass-loss rates indicates that they may not always be strong enough to be observed. The figure also shows the results for a Parker wind model with $\dot{M}=10^9$~g~s$^{-1}$ and $T=7000$~K for the same system. Decreasing the mass-loss rate by one order of magnitude in this way resulted in a spectrum with weaker spectral lines and thus fewer tracers above our threshold. At the same time, the lines probe smaller radii compared to the case with $\dot{M}=10^{10}$~g~s$^{-1}$. This behaviour is remarkably similar for different spectral lines: there is an almost monotonously increasing relation between the line-forming radii and the mass-loss rate. The same is true for the transit depth. This means that we do not identify any spectral lines that are particularly (in)sensitive to the mass-loss rate.

\subsection{Using an SED of a young star} \label{sec:young_K1}
\begin{figure}
   \centering
   \includegraphics[width=0.8\hsize]{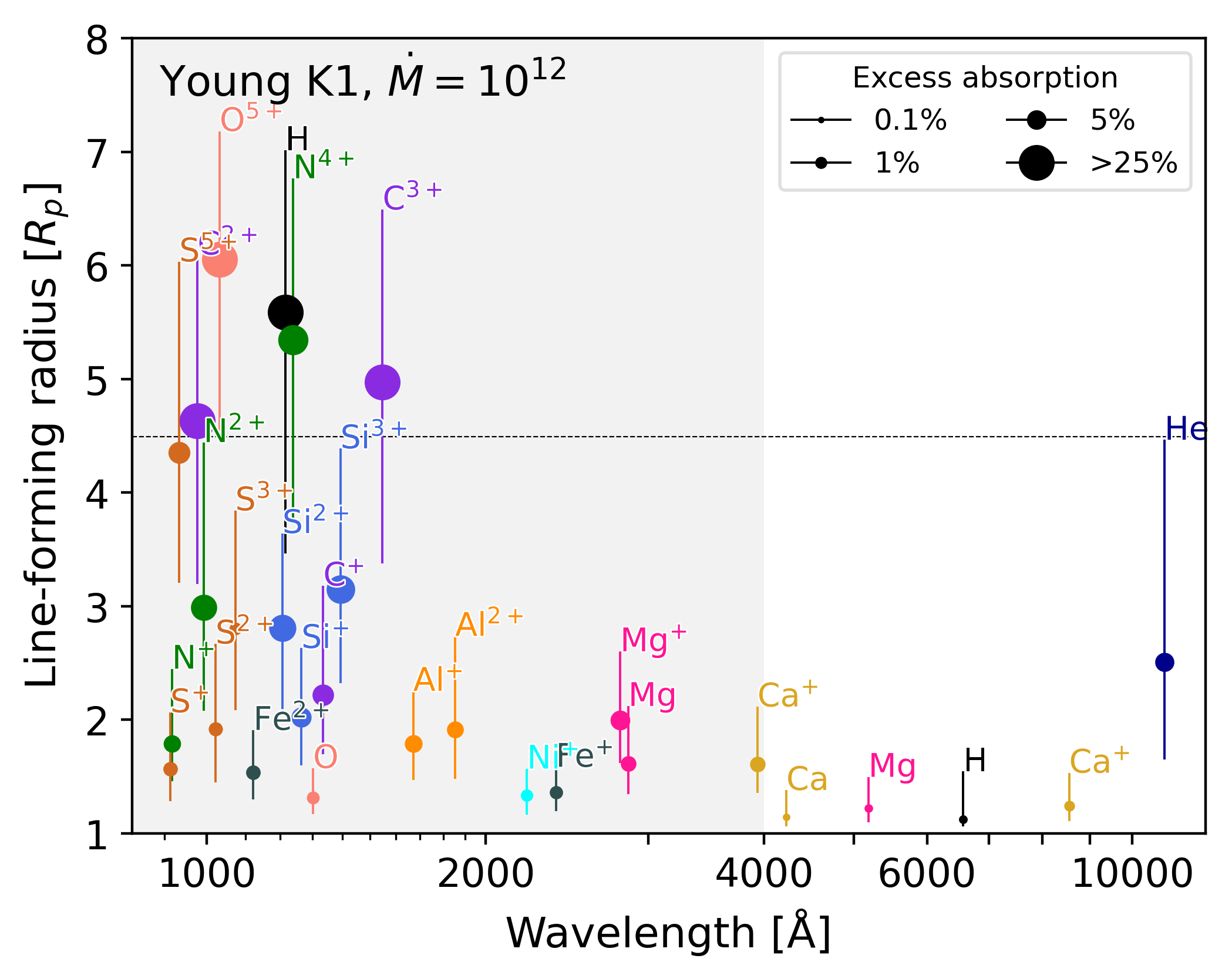}
      \caption{Line-forming radii for the hot Jupiter planet around a young K1 star, using a Parker wind model with $\dot{M}=10^{12}$~g~s$^{-1}$ and $T=12000$~K. The stellar spectrum was constructed by scaling the X-ray and EUV portions of the MUSCLES spectrum of HD~97658 to match the observed high-energy flux of V1298~Tau.}
         \label{fig:209_young_K1}
   \end{figure}

Thus far we have used SEDs of mature stars. Young stars have much higher XUV flux levels, thereby driving stronger planetary outflows through photoevaporation \citep[e.g.][]{lammer_atmopsheric_2003}. To investigate the observable signatures of such a system, we simulated the hot Jupiter planet around a young K1 type star. We took the MUSCLES spectrum of HD~97658 and multiplied the X-ray and EUV parts by factors of 2014 and 32, respectively, to match observations of the XUV flux of the young V1298~Tau system by \citet{maggio_xuv_2023} (the SED is shown in Fig. \ref{fig:SEDs}). We assumed a mass-loss rate of $\dot{M}=10^{12}$~g~s$^{-1}$ and found a self-consistent temperature of $T=12000$~K. The line-forming radii are shown in Fig. \ref{fig:209_young_K1}.

The high stellar XUV flux results in an outflow with a much higher degree of ionization compared to the mature K1 case (bottom left panel of Fig. \ref{fig:209_10}). This is reflected in the tracers we find, as the C~IV, O~VI, N~V and Si~IV are among the strongest lines in the spectrum. At the same time, it means that lines originating from atomic or singly ionized species are not much stronger than in the mature K1 case, even though the total density is higher due to the higher mass-loss rate. The high $T/\mu$ also means that the outflow velocity is higher by a factor of a few, and the spectral tracers should thus be highly broadened. At fixed mass-loss rate, such an increase in velocity would mean a decrease in density and thus generally weaker spectral lines. Therefore, the fact that we still see strong spectral lines in Fig. \ref{fig:209_young_K1} is because we used a higher mass-loss rate.

\subsection{Dependence on the computational domain and outflow geometry} \label{sec:comp_domain}
\begin{figure}
   \centering
   \includegraphics[width=\hsize]{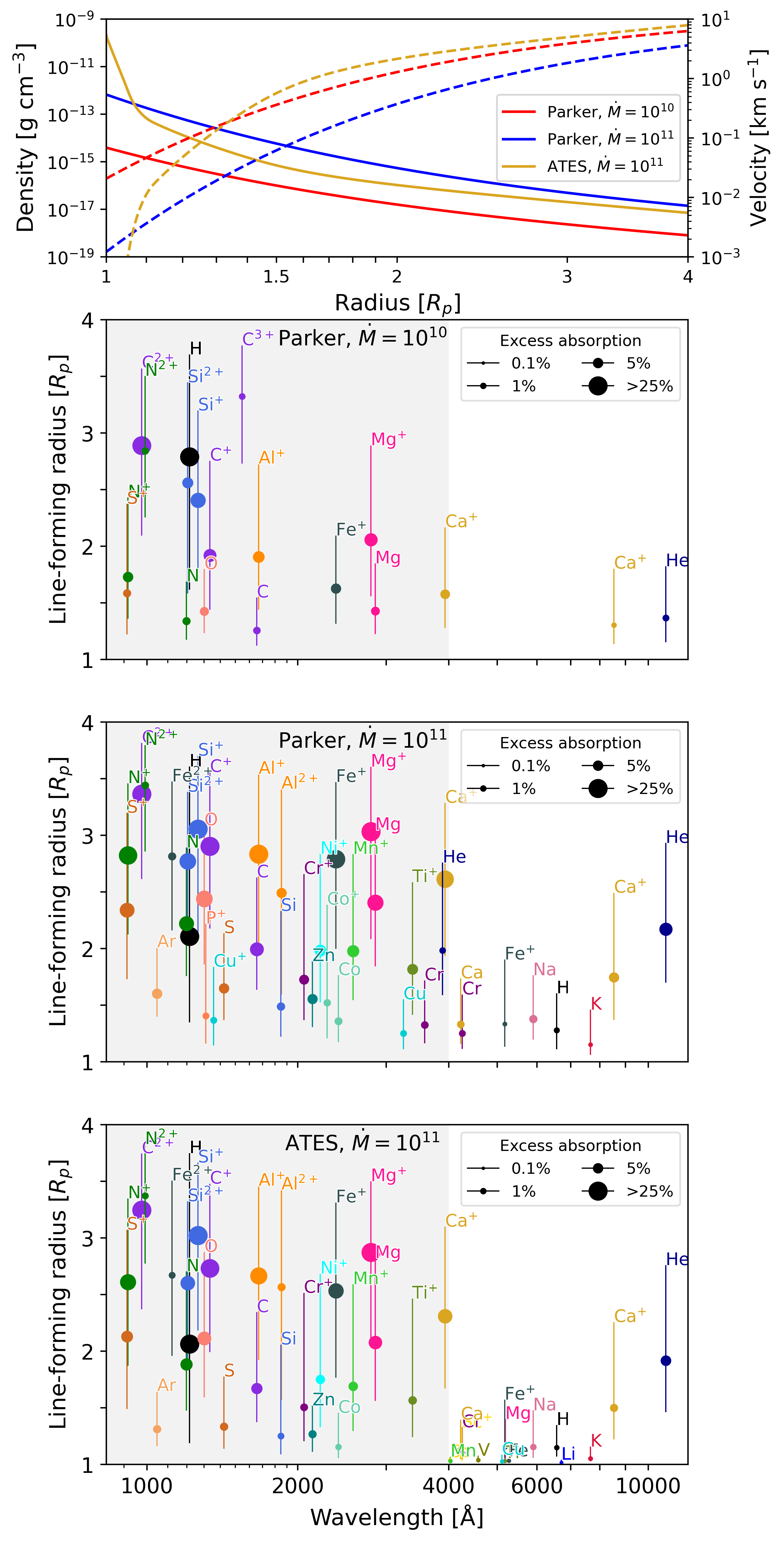}
      \caption{Line-forming radii for the hot Jupiter planet around a G2 star for different outflow profiles and with calculations truncated at 4$R_p$. The top panel shows the density and velocity profiles (dashed). The second panel shows the result for a Parker wind profile with $\dot{M}=10^{10}$~g~s$^{-1}$ and $T=8000$ (same system as in the upper right panel of Fig. \ref{fig:209_10}, but now stopped below the Hill radius). The third panel shows the result for a Parker wind profile with $\dot{M}=10^{11}$~g~s$^{-1}$ and $T=8000$. The bottom panel shows the results for ATES, which predicted a mass-loss rate of $\dot{M}=10^{11}$~g~s$^{-1}$. We find additional lines at small radii ($\lesssim 1.1R_p$) when using ATES because its density profile increases steeply in the lower atmosphere.}
         \label{fig:209_comparison_ATES}
   \end{figure}

It is important to consider the impact of the upper boundary of the computational domain for lines that form at high altitudes. Our \texttt{Cloudy} simulations extend to 8 planetary radii, even if the Hill radius of the planet is smaller. As the results show, many lines have significant contributions from material beyond the Hill sphere, where the assumption of a 1D spherically symmetric outflow is likely not valid. We chose to still include the material beyond the Hill sphere to allow for a qualitative comparison between the line-forming radii of different spectral lines, as well as between different stellar SED shapes. In Fig. \ref{fig:209_comparison_ATES} we show line-forming radii for the hot Jupiter planet around a G2 star when we stop our calculations at 4 planetary radii (a bit below the Hill radius of 4.6$R_p$). For the most part, the middle two panels of Fig. \ref{fig:209_comparison_ATES} appear as `vertically squashed' versions of the corresponding panels in Fig. \ref{fig:209_comparison_Mdot}, i.e. the lines that form significantly below the Hill radius are hardly affected, while the high-altitude lines are less strong and are now predicted to form at smaller radii. This means that for these lines, similar to the Ly-$\alpha$ line, the 3D geometry of the outflow will presumably have a strong influence on the line shape, strength, and where it forms. The 3D outflow geometry and resulting observational signatures can depend on many factors, including the stellar wind \citep{mccann_morphology_2019}, radiation pressure \citep{bourrier_3d_2013}, flares and coronal mass ejections \citep{hazra_impact_2021}, as well as the planet day- and nightside differences \citep{carroll-nellenback_hot_2017} and magnetic field \citep{khodachenko_impact_2021}. These effects may shape the outflow into highly non-spherical configurations such as an up- and downstream two-arm structure or a cometary tail. We note that the upper boundary of the atmosphere is a free parameter in our simulations, but observations of high-altitude spectral lines should be able to constrain the extent of the absorbing material, for example by measuring the time of ingress. To provide a rough estimate, the hot Jupiter planet at a distance of 0.05~AU from a G2 star has an orbital velocity of around 1$R_p$ per 0.2 hours, so that lines forming around 5$R_p$ may show significant absorption already an hour before transit. 

For our simulations of the hot Neptune planet (Fig. \ref{fig:11_10}) most lines seem to cluster around a line-forming radius of $\sim$4.5$R_p$ with large standard deviations. For this planet, the Parker wind density profiles are less `steep' than for the hot Jupiter, and thus the relative contribution of material at larger radii increases. In fact, the bulk density decreases by roughly three orders of magnitude from $1 R_p$ to $8 R_p$, but the volume of a spherical shell increases by a factor $8^3\approx500$, so that the total absorbing gas mass hardly changes as a function of radius. This makes the spectral line formation particularly sensitive to the upper boundary of the simulation, especially if it originates from a species whose relative abundance does not vary strongly with radius. The histogram in the bottom panel of Fig. \ref{fig:lfa_EW_definition} will look flat for such a spectral line and it will thus tend to form around the middle of the radial domain ($4.5R_p$ for our simulations) with little dependence on the line strength. Similarly to the high-altitude lines discussed in the paragraph above, the outflow geometry will ultimately shape the lines for this type of planet.

\subsection{Using an ATES outflow profile} \label{sec:ates}
In Sec. \ref{sec:mdot}, we investigated how the spectral tracers depend on the mass-loss rate of the Parker wind profile. Here, we study the effect of using an outflow model generated with the photoionization hydrodynamics code ATES \citep{caldiroli_irradiation-driven_2021}. The main difference with a Parker wind model is that ATES additionally solves the equation of energy conservation such that it self-consistently calculates the temperature structure instead of assuming an isothermal profile (see also Sec. 2 in \citealt{linssen_constraining_2022}). In doing so, ATES predicts the mass-loss rate instead of assuming it, but its atmospheric structure is generally similar to a Parker wind model with the same mass-loss rate and a comparable temperature (see top panel of Fig. \ref{fig:209_comparison_ATES}). ATES does however have the planet equilibrium temperature and a manually chosen density as lower boundary conditions at 1$R_p$. The equilibrium temperature is much lower than the thermospheric temperature and the density at planet radius is typically chosen as $10^{14}$~cm$^{-3}$ \citep{caldiroli_irradiation-driven_2021,caldiroli_irradiation-driven_2022}. Together, this results in a much steeper and denser lower atmospheric ($\lesssim 1.1 R_p$) density structure than that of Parker wind models. Additionally, the velocity profile of ATES tends to be steeper at smaller radii, as it transitions into a hydrostatic (i.e. velocity$\sim$0) lower atmosphere, while a Parker wind model does not necessarily do so. Precise measurements of spectral line broadening might in this way be able to distinguish between these two outflow profiles. 

We ran ATES for our hot Jupiter planet around a G2 star and post-processed the resulting density and temperature structures with \texttt{Cloudy} in order to make synthetic spectra. ATES predicts a mass-loss rate of $10^{11}$~g~s$^{-1}$ for this system. We compare the spectral tracers in Fig. \ref{fig:209_comparison_ATES}. As expected, the results are very similar. The Parker wind model predicts a few lines to be stronger at radii of roughly 1.5$R_p$ such as e.g. Cu~II~$\lambda$~1359 and Cr~I~$\lambda$~3580. The ATES model predicts a few additional lines at small radii at optical wavelengths, such as e.g. V~I~$\lambda$~4578 and Li~I~$\lambda$~6710. Both these differences can be explained by a higher density of the considered model at these radii. The sub-1.1$R_p$ radii of the extra lines found by ATES indicate that they do not really probe the escaping atmosphere, but rather the transition region from hydrostatic `lower' atmosphere to hydrodynamic `upper' atmosphere. We caution that this modeled region is somewhat uncertain; on our end because the lines form in only the few deepest computational atmospheric layers and hence the calculated radii are not very precise, but also because one has to assume and fix a density at the lower boundary of the computational domain with ATES, which allows arbitrarily scaling the strength of these spectral lines if a well-motivated or observationally constrained value is lacking. \cite{salz_simulating_2016} showed in their Fig. 4 that this `reference' density does not influence the structure of the outflow above $\gtrsim 1.1 R_p$ (for TPCI simulations, which produce similar results to ATES).

As explained in Sec. \ref{sec:modeling_with_cloudy}, we chose to assess the self-consistency of the Parker wind models in the 1-2$R_p$ region. As the results throughout this work show, we find many spectral lines that originate at radii significantly larger than this, where we in principle may deviate from the self-consistent temperature and hence density and velocity structure, as these are coupled. However, as the top panel of Fig. \ref{fig:209_comparison_ATES} shows, even at these larger radii, the structure of Parker wind models are relatively similar to ATES models of comparable mass-loss rate. We thus do not consider the fact that we have to make a choice about where to assess the self-consistency of the Parker wind models to be a major limitation of our approach.

\subsection{Dependence on the composition} \label{sec:metallicity}
In this work, we assumed a solar composition atmosphere. A change in the metallicity may have important consequences for the outflow structure and observed spectrum. The structure will change because the metallicity affects the mean molecular weight, but also because metals can be efficient coolants in the upper atmosphere \citep{fossati_non-local_2021} and hence influence the temperature structure of the outflow. The spectrum will change as a result of the different outflow profile, but also because the strength of spectral lines originating from metal species simply scales with the metal abundance, although not necessarily linearly. Especially smaller planets can be expected to have (highly) supersolar metallicities \citep[e.g.][]{fortney_framework_2013, thorngren_massmetallicity_2016, wakeford_hat-p-26b_2017} and the observed signatures may depend heavily on it. At the same time it may prove important to consider a gas composition that is not constant with radius. Heavier elements are more difficult to drag along in the planetary wind and they may preferentially reside at lower altitudes. Specific elements that are easily condensed into clouds in the lower atmosphere may also be less abundant in the upper atmosphere. We will study non-solar composition outflows in a future work.

\subsection{Targeting complementary spectral lines}
In general, observing multiple lines for the same planet will put stronger constraints on models and model parameters, as a given model then has to be able to fit these lines simultaneously. \citet{huang_hydrodynamic_2023} recently presented such a multi-line analysis for combined UV and optical observations of different escaping species. They highlighted the fact that combining several features helped to break the degeneracy between their model parameters. Some spectral lines complement each other well and are particularly useful in constraining theoretical models. The line-forming radii that we calculate in this work allow us to explore such specific synergies between different spectral tracers. For example, lines that form at similar radii should also have similar temperatures and outflow velocities, which will both contribute to spectral line broadening. Since the thermal broadening also depends on the mass of the absorbing species, observing two spectral lines from species with different masses may help to reveal the dominant broadening mechanism of the lines (in absence of other line broadening effects). Namely, such lines forming at similar radii should have similar broadening if the velocity dominates, but have different broadening due to the different mass if the temperature dominates. An example of this is the (heavy) Fe~II~$\lambda$~2382 multiplet combined with the (lighter) He~I~$\lambda$~10833 triplet or O~I~$\lambda$~1302 lines. 

It will also be useful to observe lines that form at different radii, which may help to constrain the temperature and/or velocity profile of the outflow. An example of this is combining the C~I~$\lambda$~1657, C~II~$\lambda$~1335, C~III~$\lambda$~977 and C~IV~$\lambda$~1548 lines. These lines span a large range of radii and little assumptions would have to be made about the composition of the upper atmosphere as they all originate from the same element. This series of lines would additionally allow one to study the drag of carbon to high altitudes in the outflow (which is currently included in our model through a constant composition with radius). The same could be done for a heavier species by targeting the Si~I~$\lambda$~1850, Si~II~$\lambda$~1264, Si~III~$\lambda$~1206 and Si~IV~$\lambda$~1393 series.

\section{Summary}
Escape of exoplanet atmospheres has thus far been observed in a handful of spectral lines, with the Ly-$\alpha$ line and metastable helium triplet being the most prolific. Expanding on this suite of spectral lines would help to better constrain the many different models used to interpret observations and get a more complete view of upper atmospheres. In this work, we aim to find additional spectral lines tracing upper atmospheres that could be targeted with observations. We do this by making synthetic transmission spectra of hydrodynamically escaping atmospheres and identifying the strongest spectral lines.

We describe the upper atmosphere as a solar composition 1D Parker wind and use the NLTE photoionization code \texttt{Cloudy} to obtain the radial density profiles of each absorbing species. We post-process these density profiles with the available line coefficients in the NIST database to compute the transmission spectrum from the FUV to the NIR. We make transit spectra for generic hot Jupiter and hot Neptune planets orbiting stars of spectral types ranging from A0 to M2. For every transit spectrum, we construct a sample of suitable escape tracers that includes the strongest spectral line of each atomic/ionic species if it has a transit depth stronger than 1\% or 0.1\% in the UV or VIS/NIR, respectively. In this way, we identify many spectral lines that have not yet been observed in an exoplanet atmosphere, most of which are at ultraviolet wavelengths. 

We want to know which parts of the outflow are probed by a given spectral line and thus define a line-forming radius based on the contribution of each radial atmospheric layer to the `observed' spectral line. This illustrates that the different tracers we find probe different line-forming radii, from the `lower' regions where the outflow is launched to the `upper' regions where interactions with the stellar environment are expected to shape the lines. The information of where the lines form helps to find potential synergies between them, and we give a few examples of how one might leverage observations of multiple lines to better constrain the structure of the escaping atmosphere.

\begin{acknowledgements}
    We are grateful to A. Caldiroli and the ATES team, as well as G.J. Ferland and the Cloudy team, for making their codes public and providing support. We appreciate the helpful conversations with M. MacLeod and F. Nail. We thank SURFsara (www.surfsara.nl) for the support in using the Lisa Compute Cluster. AO gratefully acknowledges support from the Dutch Research Council NWO Veni grant.
\end{acknowledgements}

\bibliographystyle{aa}
\bibliography{library}

\begin{appendix}

\section{Additional figures} \label{app:figures}

   \begin{figure}[h]
   \centering
   \includegraphics[width=\hsize]{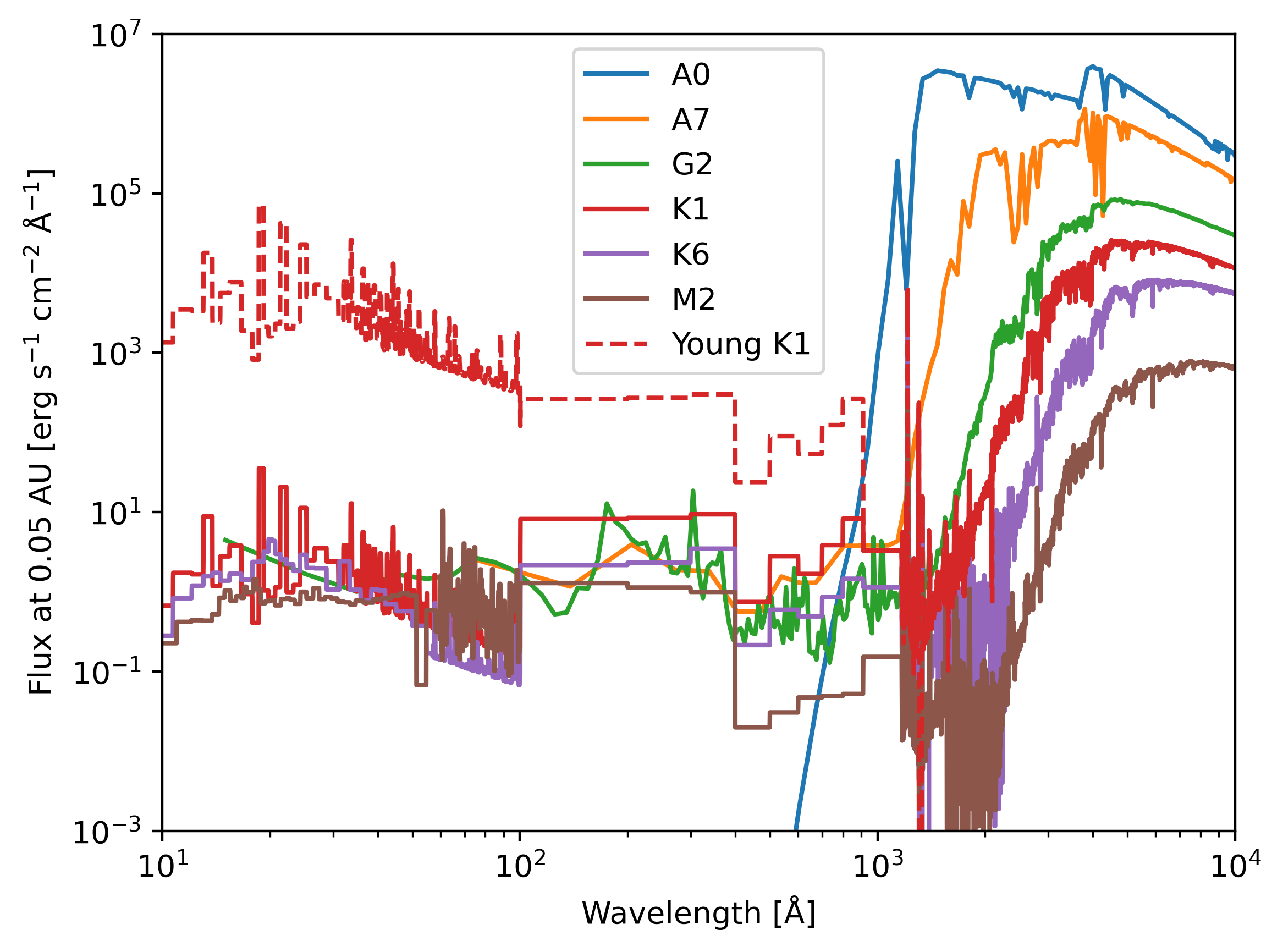}
      \caption{The stellar SEDs used throughout this work. References for these spectra are in Sec. \ref{sec:model_grid} and in Sec. \ref{sec:young_K1} for the young K1 spectrum.}
         \label{fig:SEDs}
   \end{figure}

\section{Calculating synthetic transmission spectra} \label{app:RT}
\begin{figure}
   \centering
   \includegraphics[width=\hsize]{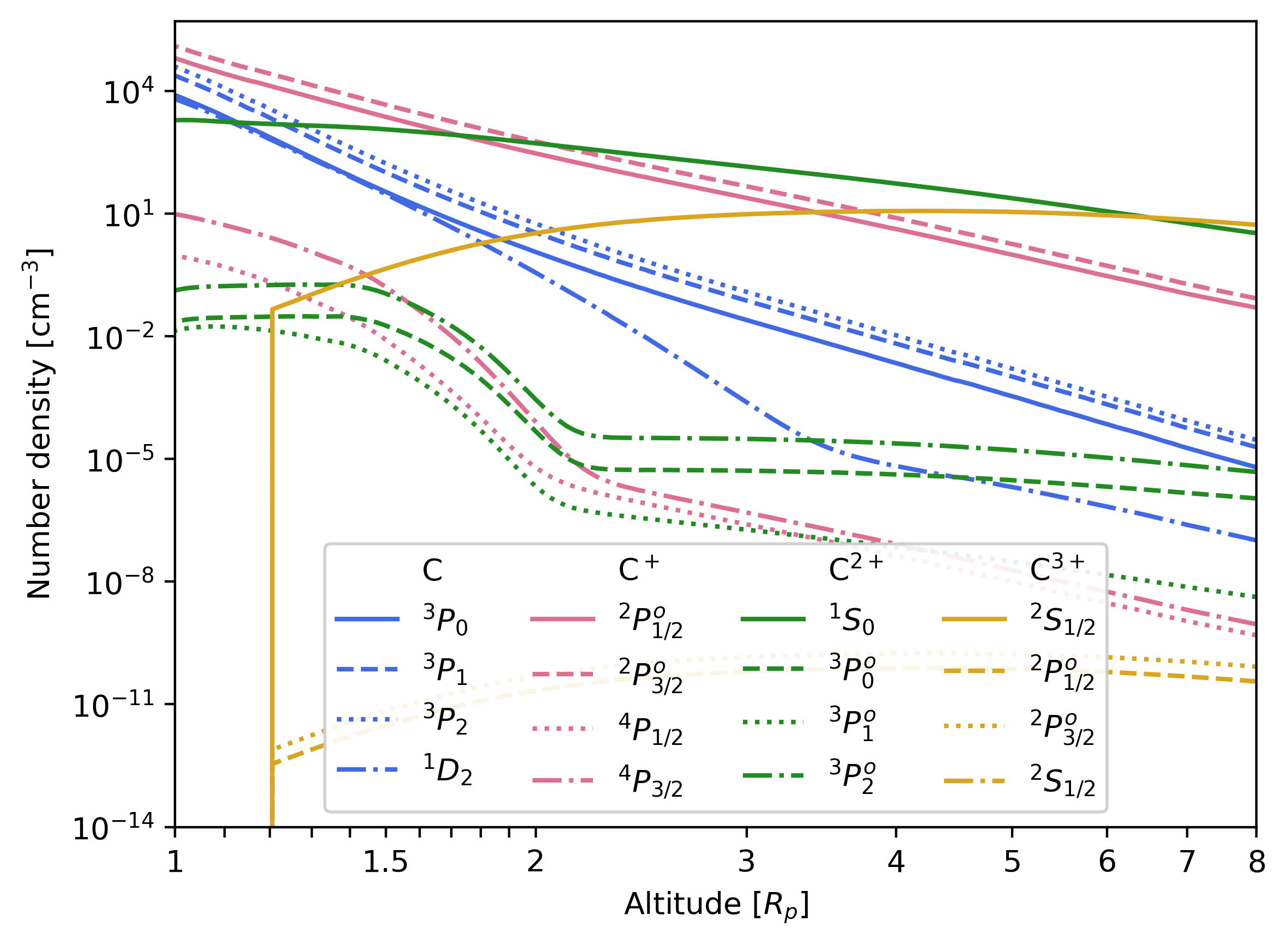}
      \caption{Density structures of different carbon ions and their energy levels as returned by \texttt{Cloudy} for the hot Jupiter planet around a G2 star. The blue, pink, green and yellow colors correspond to atomic, single, double and triple ionized carbon, respectively. For each of these, the density structures of the four lowest energy levels are shown in different line styles. We post-process such density structures with the coefficients from the NIST database to produce synthetic transmission spectra. For example, the C~I~$\lambda$~1657 multiplet is a blend of lines originating from the $^3P_0$, $^3P_1$ and $^3P_2$ levels, while the C~III~$\lambda$~977 line originates solely from the ground-state (i.e. $^1S_0$).}
         \label{fig:ion_structure}
   \end{figure}

In this work, we do not use \texttt{Cloudy}'s own predicted spectrum, as it has a spectral resolution that is too low for our purposes and furthermore does not include the effects of Doppler broadening due to the outflow velocity. Instead, we use \texttt{Cloudy} to obtain the densities of the different energy levels of all atoms and ions up to six times ionized (higher degrees of ionization are typically not reached in our simulations) as a function of radius. For most metal species, \texttt{Cloudy} resolves the lowest $\sim$15 energy levels, which is satisfactory since we are interested in strong lines that usually originate from the ground state or the lower excited states. Fig. \ref{fig:ion_structure} shows a subset of the density structures of different carbon atomic/ionic energy levels. For each of these atomic and ionic species, we download a table of spectral lines from the NIST database. For every spectral line, these tables specify the wavelength, the energy level of the atom/ion it originates from, the oscillator strength (used to calculate the cross-section) and the Einstein A coefficient (used to calculate the natural broadening). 

Ultimately, the Parker wind velocity structure, \texttt{Cloudy}'s temperature structure and density structures of the energy levels of each atomic/ionic species, together with the spectral line coefficients from the NIST database, allow us to perform radiative transfer calculations using the equations in Section 3.4 of \citet{oklopcic_new_2018}. In summary, we project the 1D density, velocity and temperature profiles onto a 2D semi-circle and calculate the optical depth $\tau$ at frequency $\nu$ along rays with different impact parameters $y$ from the planet as
\begin{equation}
    \tau_\nu(y) = \int n(x) \sigma_0 \Phi(\Delta \nu) \mathrm{d}x,
\end{equation}
where $n$ is the number density of the absorbing species, $\sigma_0$ the line cross-section, $\Phi$ the Voigt line profile and $\Delta \nu$ the offset from line center which depends on $\nu$ and the line-of-sight velocity. The extinction of each ray is $1-\mathrm{exp}(-\tau_\nu(y))$ and we find the total transit spectrum $F_{in}/F_{out}$ by adding the extinction of all rays multiplied with their projected surface on the stellar disk (which would point `out of the paper' in Fig. \ref{fig:lfa_EW_definition} so that we effectively have a 3D atmosphere). We assume a zero transit impact parameter and no stellar limb darkening.

\section{Identified lines} \label{app:lines}
Table \ref{tab:lines} lists all unique spectral lines that appear in Figs. \ref{fig:209_10} and \ref{fig:11_10} and whether they have been previously detected. Our simulations with \texttt{Cloudy} included all elements up until zinc. We post-processed these simulations to make spectra of all atoms and ions up to 6+. However, for some of these ions, the NIST database does not report any spectral lines with the coefficients needed to produce spectra. Thus, we were not able to produce spectra for the following atomic and ionic species: F, Ti$^{4+}$, Ti$^{6+}$, V$^{3+}$, V$^{4+}$, V$^{5+}$, V$^{6+}$, Cr$^{2+}$, Cr$^{3+}$, Cr$^{4+}$, Cr$^{5+}$, Cr$^{6+}$, Mn$^{2+}$, Mn$^{3+}$, Mn$^{4+}$, Mn$^{5+}$, Mn$^{6+}$, Fe$^{3+}$, Fe$^{5+}$, Co$^{3+}$, Co$^{4+}$, Co$^{5+}$, Co$^{6+}$, Ni$^{3+}$, Ni$^{5+}$, Ni$^{6+}$, Cu$^{2+}$, Cu$^{3+}$, Cu$^{4+}$, Cu$^{5+}$, Cu$^{6+}$, Zn$^{+}$, Zn$^{2+}$, Zn$^{3+}$, Zn$^{4+}$, Zn$^{5+}$, Zn$^{6+}$.

\begin{table}[h]
\caption{All unique identified spectral lines tracing upper atmospheres of Figs. \ref{fig:209_10}, \ref{fig:11_10}, \ref{fig:209_comparison_Mdot} and \ref{fig:209_young_K1} (thus excluding the additional lines found in Fig. \ref{fig:209_comparison_ATES} when using an ATES model). We give only one example reference for lines that have been previously detected.} 
\label{tab:lines}
\centering
\begin{tabular}{l l l}
\hline\hline
Wavelength & Species & Previously detected\\
(vac) (Å) & & \\
\hline                    
912.7 & S II & No\\
916.7 & N II & No\\
933.4 & S VI & No\\
977.0 & C III & No\\
991.6 & N III & No\\
1021.3 & S III & No\\
1031.9 & O VI & No\\
1048.2 & Ar I & No\\
1073.0 & S IV & No\\
1122.5 & Fe III & No\\
1199.5 & N I & No\\
1206.5 & Si III & e.g. \citet{linsky_observations_2010}\tablefootmark{1}\\
1215.7 & H I & e.g. \citet{vidal-madjar_extended_2003}\\
1238.8 & N V & No\\
1264.7 & Si II & No\\
1302.2 & O I & e.g. \citet{vidal-madjar_detection_2004} \\
1310.7 & P II & No\\
1335.7 & C II & e.g. \citet{vidal-madjar_detection_2004} \\
1358.8 & Cu II & No\\
1393.8 & Si IV & e.g. \citet{schlawin_exoplanetary_2010}\tablefootmark{2}\\
1425.0 & S I & No\\
1548.2 & C IV & No\\
1657.0 & C I & No\\
1670.8 & Al II & No\\
1850.7 & Si I & No\\
1854.7 & Al III & No\\
2056.3 & Cr II & No\\
2139.2 & Zn I & No\\
2217.2 & Ni II & No\\
2276.2 & Ca I & No\\
2286.9 & Co II & No\\
2382.8 & Fe II & e.g. \citet{sing_hubble_2019}\\
2408.0 & Co I & No\\
2576.9 & Mn II & No\\
2796.4 & Mg II & e.g. \citet{fossati_metals_2010}\\
2853.0 & Mg I & e.g. \citet{vidal-madjar_magnesium_2013}\tablefootmark{3}\\
3248.5 & Cu I & No\\
3384.7 & Ti II & No\\
3579.7 & Cr I & No\\
3889.7 & He I & No\\
3934.8 & Ca II & e.g. \citet{yan_ionized_2019}\\
\hline
4227.9 & Ca I & No\\
4255.5 & Cr I & No\\
5170.5 & Fe II & e.g. \citet{casasayas-barris_atmospheric_2019}\\ 
5185.0 & Mg I & e.g. \citet{zhang_transmission_2022}\\ 
5891.6 & Na I & e.g. \citet{charbonneau_detection_2002}\\
6564.5 & H I & e.g. \citet{jensen_detection_2012}\\
7667.0 & K I & e.g. \citet{sing_gran_2011}\\
8544.4 & Ca II & e.g. \citet{casasayas-barris_atmospheric_2019}\\
10833.3 & He I & e.g. \citet{spake_helium_2018}\\
\hline
\end{tabular}
\tablefoot{
\tablefoottext{1}{Contested by \citet{ballester_re-visit_2015}}
\tablefoottext{2}{Tentative detection}
\tablefoottext{3}{Contested by \citet{cubillos_near-ultraviolet_2020}}
}

\end{table}

\end{appendix}

\end{document}